\documentclass[runningheads]{llncs}
\usepackage[T1]{fontenc}
\usepackage{graphicx}
\usepackage{glossaries}
\usepackage{csquotes}
\usepackage{tikz}
\usepackage{float}
\usepackage{todonotes}
\usepackage{hyperref}
\hypersetup{hidelinks,
	backref=true,
	pagebackref=true,
	hyperindex=true,
	breaklinks=true,
	colorlinks=true,
	linkcolor=black!60,
	citecolor=black!50,
	urlcolor=blue,
	bookmarks=true,
	bookmarksopen=true
}

\newacronym{ids}{IDSs}{intrusion detection systems}
\newacronym{it}{IT}{information technology}
\newacronym{ml}{ML}{Machine Learning}
\newacronym{ot}{OT}{operational technology}
\newacronym{dpi}{DPI}{deep packet inspection}
\newacronym{ics}{ICSs}{industrial control systems}
\newacronym{nids}{NIDS}{network intrusion detection systems}
\newacronym{hids}{HIDS}{host intrusion detection systems}
\newacronym{ips}{IPS}{intrusion prevention systems}
\newacronym{swat}{SWaT}{Secure Water Treatment}
\newacronym{hmi}{HMI}{human machine interfaces}
\newacronym{plc}{PLC}{programmable logic controllers}
\newacronym{scada}{SCADA}{supervisory control and data acquisition}
\newacronym{apt}{APT}{advanced persistent threats}
\newacronym{ddos}{DDoS}{distributed denial of service}
\newacronym{cps}{CPSs}{cyber-physical systems}
\newacronym{cve}{CVEs}{common vulnerabilities and exposures}
\newacronym{cip}{CIP}{common industrial protocol}
\newacronym{sis}{SISs}{safety instrumented systems}
\newacronym{dcs}{DCSs}{distributed control systems}
\newacronym{dmz}{DMZ}{demilitarized zone}
\newacronym{siem}{SIEM}{security information and event management}
\newacronym{soar}{SOAR}{security orchestration, automation, and response}
\newacronym{pca}{PCA}{principal component analysis}
\newacronym{rtu}{RTUs}{remote terminal units}
\newacronym{smote}{SMOTE}{synthetic minority over-sampling technique}
\newacronym{adasyn}{ADASYN}{adaptive synthetic sampling approach for imbalanced learning}

\begin{document}

\title{Enhanced Anomaly Detection  in  Industrial Control Systems aided by Machine Learning
\thanks{The full version of the work can be found at \url{https://bora.uib.no/bora-xmlui/handle/11250/3140895}}
}
\author{Vegard Berge, Chunlei Li \orcidIMG
\\	
	Emails: \href{mailto:vegard.berge8@gmail.com}{vegard.berge8@gmail.com}, \href{mailto:chunlei.li@uib.no}{chunlei.li@uib.no}	
}
	\institute{Department of Informatics, University of Bergen, Bergen, 5020, Norway}

%
\maketitle

\begin{abstract}
Traditional 
\gls{ids} often rely on either network traffic or process data, but this single-source approach may miss complex attack patterns that span multiple layers within \gls{ics} or persistent threats that target different layers of \gls{ot} systems. This study investigates whether combining both network and process data can improve attack detection in \gls{ics} environments. Leveraging the \gls{swat} dataset, we evaluate various machine learning models on individual and combined data sources. Our findings suggest that integrating network traffic with operational process data can enhance detection capabilities, evidenced by improved recall rates for cyber attack classification. Serving as a proof-of-concept within a limited testing environment, this research explores the feasibility of advancing intrusion detection through a multi-source data approach in \gls{ics}. Although the results are promising, they are preliminary and highlight the need for further studies across diverse datasets and refined methodologies.

\end{abstract}

\section{Introduction}

As industrial environments grow increasingly interconnected, with the proliferation of technologies such as the industrial Internet of Things (IIoT) and Industry 4.0, the security of \gls{ics} becomes ever more critical. ICSs are the backbone of critical infrastructure, overseeing and automating processes in sectors such as energy, water treatment, manufacturing, and transportation. The convergence of information technology (IT) and operational technology (OT) in these systems has improved efficiency but also exposed \gls{ics} to a wide range of cyber threats \cite{ackerman2021industrial,stouffer2022guide}.  Cyber attacks targeting ICSs can have catastrophic consequences, ranging from operational disruption to physical damage and even loss of life, as many \gls{ics} components are responsible for real-time control of physical processes.
Most conventional IDSs focus on IT environments, emphasising data confidentiality. However, in the OT context of ICS, availability is the paramount concern, as systems must remain continuously operational to avoid disruption of critical services \cite{cyber_report_2023}. This shift in priorities, coupled with the unique operational characteristics of ICS, necessitates the development of specialised IDSs tailored to detect both network and process anomalies.

\subsection{Research Motivation and Problems}

The integration of network traffic data with operational process data is a promising approach to improving the detection capabilities of IDSs in ICS environments. While network-based IDSs (NIDSs) monitor network traffic for suspicious patterns, they often miss anomalies within the physical processes themselves. Conversely, process-based IDSs focus on deviations in the behaviour of the physical components, such as programmable logic controllers (PLCs) and sensors, but may lack the broader context provided by network data.
This study explores the potential for enhanced detection capabilities by incorporating both OT and IT data. This preliminary approach is not intended to replace existing detection frameworks but rather to assess the viability of enriching them with complementary data sources. Specifically, this research aims to answer the following questions: 
%

\begin{itemize}
    \item How can machine learning models be applied to detect anomalies across both network and process data in ICSs?
    \item Does the integration of these two data types improve the overall detection capabilities and reduce false positives compared to models trained on network or process data alone?
    \item What are the limitations of current intrusion detection approaches, and how can this research contribute to overcoming them?
\end{itemize}

\subsection{Main Contributions}

This research tries to contribute to the field of ICS security in these aspects:
\begin{itemize}
    \item Integrated Anomaly Detection: We propose an approach that combines network traffic data and process data to improve the detection of cyber attacks in ICS environments. This hybrid approach enables a more comprehensive detection strategy by capturing both network anomalies and operational deviations.
    \item Evaluation on Real-World Data: Using the Secure Water Treatment (SWaT) testbed dataset from iTrust Labs, this study evaluates machine learning models on realistic ICS data. The SWaT dataset includes both network and process data, providing a rich environment for testing IDSs approaches.
    \item Machine Learning Models for ICSs: A range of machine learning models, including Decision Trees, Random Forests, Support Vector Machines (SVM), and Neural Networks are applied. This research explores how these models perform binary classification, trained on a combined dataset compared to when they are trained on either network or process data alone.
    \item Actionable Insights: By evaluating the results from our models, on the variation of datasets, we try to explain what benefits might come from viewing all assets in an ICS as potential \textit{sensor} or \textit{agent} for anomaly detection.
\end{itemize}


The remainder of this paper is organised as follows: In Section 2, we start with an overview of \gls{ics}, discuss their key components and security challenges, then review related work in the field of IDSs for ICSs. In Section 3 we introduce the SWaT testbed and describe its relevance to the research. Section 4 outlines the methodology used in this study, including data preprocessing and model training. Section 5 presents the experimental results and discusses their implications. Finally, Section 6 concludes the paper and suggests directions for future research.

\section{Preliminaries}
\subsection{Overview of ICSs}

\gls{ics} integrate hardware, software, and network systems to monitor, control, and automate physical processes. These systems are essential for managing large-scale industrial operations that require precise and continuous control, where downtime can have severe economic, environmental or safety impacts \cite{stouffer2022guide}.

\gls{ics} consist of several core components that interact to ensure efficient, reliable, and safe industrial operations. Some of these components include:
\begin{itemize}
    \item \textbf{Programmable Logic Controllers (PLCs)}: \gls{plc}s are industrial computers designed to execute control tasks based on real-time input from sensors. They play an role in automating processes, such as opening valves, starting pumps.. 
    
    \item \textbf{Supervisory Control and Data Acquisition (SCADA)}: SCADA systems provide a supervisory layer that enables operators to monitor and control industrial processes remotely \cite{galloway2012introduction}. 
    
    \item \textbf{Human-Machine Interface (HMI)}: \gls{hmi} systems serve as the primary interface between operators and \gls{ics}. 

    \item \textbf{Remote Terminal Units (RTUs)}: \gls{rtu} serve as field devices that communicate with \gls{scada} systems and \gls{plc}s to collect and send process data from remote locations \cite{sullivan2016components}. 
\end{itemize}

\subsection{Cybersecurity Challenges in ICSs}

With the rise of Industry 4.0 and the Industrial Internet of Things (IIoT), \gls{ics} are now increasingly interconnected with enterprise IT networks, exposing them to cyber threats that were not considered during their initial design. This convergence of \gls{it} and \gls{ot} brings both advantages in terms of improved monitoring and control but also introduces new cybersecurity challenges \cite{ackerman2021industrial}.
Unlike traditional IT systems, where \textit{confidentiality} of data is the top priority, the most critical aspect of \gls{ics} is \textit{availability}, followed closely by \textit{integrity} \cite{cyber_report_2023}. Interruptions to control systems can lead to physical damage, safety incidents, or production halts, which makes continuous and reliable operation paramount.
There are several key challenges in securing \gls{ics}:
\begin{itemize}
    \item \textbf{Legacy Systems}: Many \gls{ics} are built on legacy hardware and software that were not designed with cybersecurity in mind. These systems often lack the ability to receive security updates, leaving them vulnerable to exploits.
    
    \item \textbf{Proprietary Protocols}: \gls{ics} frequently rely on proprietary communication protocols, such as Modbus, DNP3, and \gls{cip}. These protocols may lack encryption or strong authentication mechanisms, making them targets for cyber attacks.
    
    \item \textbf{Lack of Patching and Updates}: 
    ICSs are often not updated as frequently due to concerns about disrupting critical processes. This leads to prolonged exposure to potential cyber threats.
    
    \item \textbf{Increased Connectivity}: The convergence of \gls{it} and \gls{ot} has increased connectivity between industrial control networks and external systems, which introduces a wider attack surface for cyber adversaries. As more devices are connected to the Internet or enterprise networks, they become susceptible to malware, ransomware, and other external attacks.
    
    \item \textbf{Insider Threats}: In addition to external attacks, \gls{ics} are also vulnerable to insider threats, where employees or contractors with authorised access may misuse their privileges to compromise systems. This makes robust monitoring and intrusion detection crucial \cite{stouffer2022guide}.
\end{itemize}

\subsection{Multi-layer Security: the Purdue Model} \label{purdue_model}

To address these challenges, \gls{ics} security follows the Purdue Enterprise Reference Architecture (PERA), commonly referred to as the Purdue Model \cite{ackerman2021industrial}. The model divides \gls{ics} into different layers, each representing different operational zones within an industrial system:
\begin{itemize}
    \item \textbf{Level 0-1: Process Control}: These levels include the physical devices and sensors that interact directly with industrial processes, such as actuators, sensors, and \gls{plc}s. Security at this level focuses on protecting the real-time control of physical processes.
    
    \item \textbf{Level 2: Supervisory Control}: This level includes \gls{hmi} and SCADA systems, which provide operational oversight and process visualisation. Security concerns here include preventing unauthorised control and ensuring accurate reporting of process data.
    
    \item \textbf{Level 3: Operations Management}: At this level, the focus shifts to production management systems and databases that track operational performance. Security at this level is concerned with protecting sensitive operational data and preventing disruptions to production schedules.
    
    \item \textbf{Level 3.5: DMZ}: This level forms the boundary between the enterprise IT network and the \gls{ot} network, often containing firewalls, \gls{ids}, and other security mechanisms that help segregate the two environments.
    
\end{itemize}
By segmenting networks and applying security controls at each layer, the Purdue Model enhances the ability of \gls{ics} to withstand and contain cyber attacks. Each layer is protected individually, ensuring that if one layer is compromised, the attack cannot easily spread to other parts of the system. 



\subsection{Intrusion Detection aided by Machine Learning}

\gls{ids}s have long been a central part of cybersecurity countermeasures, particularly in information technology environments. 
\gls{ml} has become a central focus in enhancing the detection capabilities of \glspl{ids}, both in traditional \gls{it} systems and, increasingly, in \gls{ics} environments. Techniques such as \textit{Decision Trees} \cite{quinlan1993c45}, \textit{Random Forests} \cite{breiman2001random}, \textit{Support Vector Machines} \cite{cortes1995svm}, and \textit{Neural Networks} \cite{GAMAGE2020102767} have been widely applied to anomaly detection in network traffic. 
In \gls{it} environments, these models have shown considerable promise in detecting intrusions, malware, and other cyber threats, especially in scenarios where traditional signature-based systems fail.
In \gls{ics}, these techniques must be adapted to handle the operational constraints and real-time requirements specific to industrial systems, as well as the distinct characteristics of the process data.
The complexity of modern \gls{ics} environments, characterised by heterogeneous networks, proprietary protocols, and a mix of legacy and modern equipment, requires customised \glspl{ids} that can integrate both the network and the processing of data to detect cyberphysical threats \cite{galloway2012introduction}. 

Some research efforts have been made on ML-aided intrusion detection for ICS environments. Research by Goh et al. \cite{goh2016} developed a deep learning-based \gls{ids} specifically for \gls{ics}, leveraging process data to detect anomalies. Their work highlights the potential for deep learning to improve detection accuracy by modelling the complex relationships between sensors, actuators, and control systems. Similarly, Kravchik and Shabtai \cite{kravchik2018} employed convolutional neural networks (CNNs) to detect cyber attacks in \gls{ics}, demonstrating how \gls{ml} techniques can process high-dimensional time-series data to uncover subtle anomalies.
There is also an increasing focus on hybrid approaches that combine both network and operational data. Studies have shown that combining network-based and process-based anomaly detection can significantly improve overall security posture by identifying attacks that target both network infrastructure and physical processes managed by \gls{ics} components \cite{jadidi2022automated}.
Konijn \cite{konijn2022multi} presented a comprehensive study that integrates network traffic and process data to enhance \gls{ids} capabilities in industrial environments. This multi-domain approach aims to detect complex, multi-stage cyber attacks that affect both network communications and physical processes. Konijn highlighted that much of the literature in \gls{ics} security focuses either on network or physical sensor data, with few studies investigating their integration. This gap is particularly critical for detecting sophisticated attacks, such as advanced persistent threats (APTs), which often span multiple layers of the Purdue model. Shalyga et al. \cite{shalyga2018anomaly} proposed an anomaly detection system that integrates network traffic and operational data using neural networks to detect cyber attacks targeting \gls{ics}. This multilayered approach is effective in detecting attacks that traditional \gls{nids} might miss, such as those that directly manipulate process variables.
Similarly, Jadidi et al. \cite{jadidi2022automated} introduced a layered \gls{ids} designed to detect flooding attacks in \gls{ics} environments. Their system combines network traffic analysis with physical sensor data, demonstrating the importance of multi-domain analysis for comprehensive intrusion detection. The inclusion of both data sources allows for the detection of attacks that would otherwise be missed if only one data type were considered.

%
%
%

\smallskip

While significant advancements have been made in \gls{ids} for \gls{ics}, the integration of \gls{ml} techniques and multi-domain data remains a critical area for further research. This work seeks to contribute to the field by combining network and process data to enhance anomaly detection capabilities, building on the foundation established by earlier studies.

\section{SWaT Testbed} \label{section_swat_testbed}

The testbed \gls{swat}, developed by iTrust Labs at Singapore University of Technology and Design (SUTD), is a high-fidelity ICS testbed that simulates the operation of a modern water treatment facility \cite{swat_technical_report}. Commissioned in 2015 with support from the Ministry of Defence (MINDEF) of Singapore, the testbed is designed to provide researchers with a realistic environment to study cyber-physical security threats in \gls{ics} environments. The SWaT testbed is widely used in academic and industrial research, particularly in the areas of anomaly detection and other cybersecurity solutions.

The testbed spans approximately 90 square metres and consists of six distinct stages that collectively simulate a real-world water treatment process, from raw water intake to reverse osmosis and membrane cleaning. Each stage is controlled by Programmable Logic Controllers (\glspl{plc}) and other industrial grade devices commonly found in \gls{ot} environments \cite{swat_technical_report}. This infrastructure closely mirrors that of modern water treatment plants, making it an ideal platform for studying the impact of cyber attacks on both physical processes and network infrastructure.

\subsection{Architecture of the SWaT Testbed}

\begin{figure}[t]
	\centering
	\includegraphics[width=\textwidth]{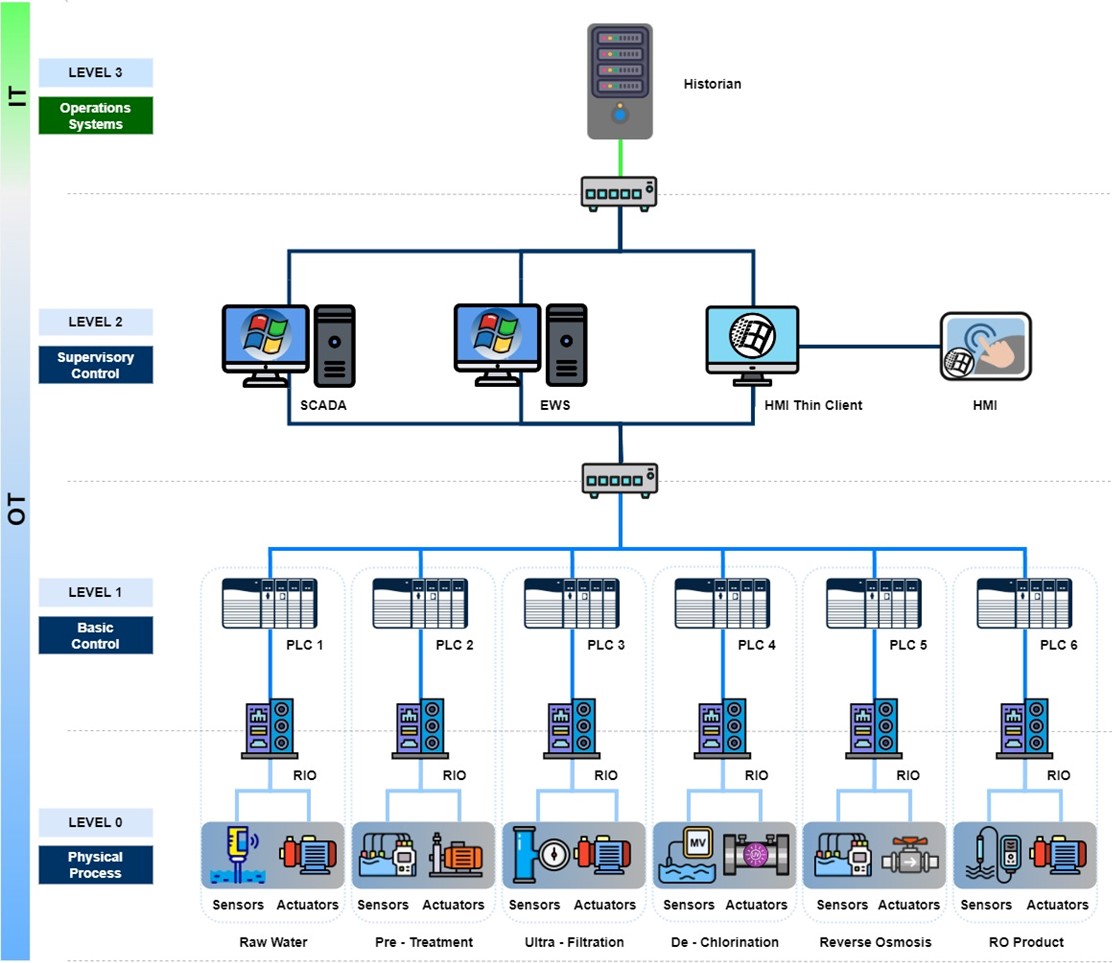}
	\caption{Network Architecture of the SWaT Physical Process \cite{swat_technical_report}}
	\label{fig:swat_block_diagram}
\end{figure}

The SWaT testbed is designed around six sequential stages of water treatment:

\begin{itemize}
    \item \textbf{P1: Raw Water Supply} - pumping raw water into the system.
    \item \textbf{P2: Chemical Dosing} - adds chemicals for purification and water safety.
    \item \textbf{P3: Ultrafiltration} - filtering out particles from the water.
    \item \textbf{P4: De-chlorination via Ultraviolet (UV) Light} -  using UV light to remove chlorine and other contaminants.
    \item \textbf{P5: Reverse Osmosis} - purifying water by filtering it through membranes.
    \item \textbf{P6: Backwash Cleaning} - cleaning the ultrafiltration membranes to prevent fouling.
\end{itemize}
Each stage is equipped with its own sensors, actuators, and \gls{plc} to control the operation of the physical processes. The control system in the SWaT testbed mimics those in real industrial environments, where physical processes are closely monitored and controlled by a network of \gls{plc}, \gls{hmi}, and \gls{scada} systems \cite{swat_technical_report}. In addition, the SWaT testbed is segmented according to the Purdue Model, with different network zones assigned to different functional areas, such as the plant floor, the control room, and the enterprise network.
\begin{figure}[h]
	\centering
	\includegraphics[width=\textwidth]{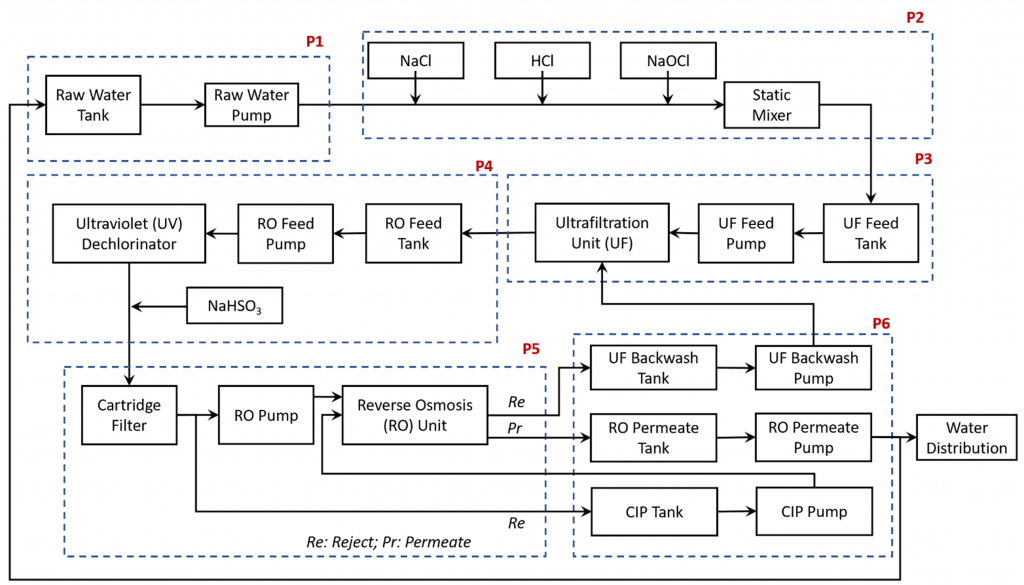}
	\caption{Block Diagram of the SWaT Physical Process \cite{swat_technical_report}}
	\label{fig:swat_block_diagram}
\end{figure}

\subsection{Cybersecurity Research on the SWaT Testbed}

The SWaT testbed has been used in numerous cybersecurity studies to evaluate the effectiveness of \gls{ids}, \gls{ips}, and other defensive mechanisms in \gls{ics}. By providing real-time access to both network traffic and physical process data, the testbed allows researchers to simulate and study the impact of various cyber attacks on critical infrastructure.

The SWaT dataset, publicly available through iTrust Labs, contains both normal operation data and data collected during a series of staged cyber attacks on the water treatment process. These attacks include manipulation of \glspl{plc}, sensor spoofing, and other tactics that affect both the physical and network layers of the system. The dataset includes logs of sensor readings, actuator states, and network traffic, making it an ideal resource for the development of machine learning-based \glspl{ids} \cite{swat_technical_report}. 

Researchers have used the SWaT dataset to study a wide range of attack scenarios, including insider threats, network-based attacks, and attacks on physical processes. One of the key strengths of the SWaT testbed is its ability to provide a comprehensive view of both the network and physical layers, enabling a deeper understanding of how cyber attacks propagate through \gls{ics} environments and affect physical processes.


In this study, the SWaT dataset is used to evaluate the performance of various machine learning models for anomaly detection in \gls{ics}. Using both network traffic data and physical process data, this research aims to investigate the potential benefits in detecting attacks at various levels in \gls{ics} that may go undetected by more traditional \gls{it} based \gls{ids}. The data set provides a unique opportunity to explore the interplay between network and process anomalies, offering insight into how these systems can be protected, as well as providing valuable insight from cyber physical threats.

\section{Methodology}
This section outlines the dataset, preprocessing steps, machine learning models, feature extraction techniques, and the evaluation metrics used in the study.

\subsection{A6 - dataset}
The dataset used for this research, the A6 dataset from iTrust, is derived from the \gls{swat} testbed, described in Section \ref{section_swat_testbed}. Most importantly, this dataset consists of two main data streams:
\begin{enumerate}
    \item \textbf{Network traffic data}: Captures communication between \gls{plc} and other devices, including network logs, packet details, and communication protocols. In terms of the Purdue Model \ref{purdue_model}, this would be data laying on layer 1 through 3 (and 3.5).  
    \item \textbf{Process data}: Includes sensor readings and actuator states from the various stages of the water treatment process. These data points are critical for monitoring the physical behaviour of the system. From the Purdue Model \ref{purdue_model}, this would be items at layer 0. 
\end{enumerate}
The dataset comprises normal operation data, as well as data collected during a series of staged cyber attacks, which target both the network and physical layers of the system \cite{swat_technical_report}. The attacks include various forms of network intrusions, sensor spoofing, and manipulation of actuators, simulating real-world cyber-physical threats. This dual-domain dataset is ideal for testing intrusion detection systems that can monitor both network and process anomalies.

\subsection{Network Data Extraction with Zeek} \label{subsec_zeek_extraction}

Zeek is an open-source tool designed for network traffic analysis, capable of dissecting network packets and producing structured logs that provide detailed insights into the behaviour of communication protocols and network traffic patterns. Zeek's flexibility in supporting custom log generation allowed us to tailor the analysis to focus on the key communication logs relevant to the \gls{swat} testbed experiment.

In this work, we focused on three specific types of logs generated by Zeek: \texttt{cip.log}, \texttt{conn.log}, and process variable communication. These logs were selected based on their relevance to different layers of the Purdue Model, ensuring a comprehensive analysis of the interactions.
\begin{itemize}
    \item \textbf{conn.log}: The \texttt{conn.log} provides a general overview of the network connections in the system, recording details such as source and destination IP addresses, protocol types, connection durations, and byte counts. This log gives insight into the overall network traffic, corresponding to Purdue Level 3. It is particularly useful for identifying network-level anomalies, such as unauthorised connections or unusual traffic patterns that might indicate the presence of a network-based attack.
    \item \textbf{cip.log}: This log captures communication using the \gls{cip}, a widely used protocol for communication between \gls{plc} and other industrial devices in the \gls{ot} environment. By analysing \texttt{cip.log}, we gain visibility into the communication between critical industrial components (Purdue Level 1 and 2) and can detect deviations from normal protocol behaviour.
    \item \textbf{Process variable communication}: Process data, collected from sensors and actuators, corresponds to Purdue Level 0 (sensors and actuators). In this research, process data was collected alongside network data, allowing us to monitor the behaviour of physical processes. This dual-layer approach enables the detection of anomalies that may not be visible through network traffic alone, such as sensor spoofing or manipulation of actuators.
\end{itemize}
The combination of these three data sources allows us to map different layers of the Purdue Model to the corresponding logs, creating a comprehensive view of both network-level and process-level interactions. 
%
%
%
The choice of focussing on \texttt{conn.log}, \texttt{cip.log}, and process variable communication was a deliberate decision aligned with the hierarchical structure of the Purdue model \ref{purdue_model} as well as the availability of data in our dataset. By having logs that reflects both the high-level network communications to low-level physical process variables we aim to capture new aspects and correlation from the \gls{ics}. This layered approach mirrors the multilayered security strategy advocated by industrial security frameworks, such as ISA/IEC 62443 \cite{iec62443}, and highlights the importance of monitoring all layers of \gls{ics} to detect anomalies in the physical and network processes.

\subsection{Feature Extraction from Zeek Logs}

Once the Zeek logs were generated, the next step is to extract meaningful features for machine learning models. Feature extraction from the \texttt{conn.log} and \texttt{cip.log} files focused on identifying key attributes related to the communication flow between components and identifying anomalous behaviours indicative of cyber attacks. Key features extracted from these logs include:
\begin{itemize}
    \item \textbf{\texttt{conn.log}}: Source and destination IPs, port numbers, duration of connections, protocols used, packet sizes, and connection status. These features help in identifying unauthorised connections, port scanning activities, or unusual traffic patterns that may suggest a Distributed Denial of Service (DDoS) attack or a network intrusion.
    \item \textbf{\texttt{cip.log}}: Specific attributes related to \gls{cip} communication, such as command types, response codes, and message sizes. These features are critical for detecting anomalies in the control communication between \glspl{plc} and sensors or actuators, which could indicate a manipulation of industrial processes.
    \item \textbf{Process Variable Communication}: Sensor readings (e.g., flow rates, pressure, valve states) were treated as time series data. Key features included average values, standard deviations, and abrupt changes or spikes that could indicate a manipulation of the physical system. This layer provides insight into whether the observed network anomalies correspond to actual changes in the physical process.
\end{itemize}
These extracted features formed the basis of our machine learning models, providing a combination of network-level and process-level data that enables the detection of both traditional IT-based attacks (e.g., unauthorised access, DDoS) and OT-based attacks (e.g., sensor spoofing, actuator manipulation).

\subsection{Data Preprocessing}
Before training the \gls{ml} models, several preprocessing steps were applied to ensure the quality and usability of the dataset:

\begin{itemize}
    \item \textbf{Handling Missing Data}: Missing or corrupted data points in both network traffic and process logs were handled through \textit{interpolation} where possible. 
    \item \textbf{Data Labelling}: The dataset contains semi-labelled data, with timestamps indicating known attack windows. Data within these windows were labelled as ``malicious'', while data outside were considered ``benign''. Given the lack of precise attack indicators in the dataset, this labelling was manually aligned with the attack events documented in the SWaT testbed timeline \cite{swat_technical_report}.
    \item \textbf{Normalisation}: To ensure that all features were on a comparable scale, especially for models sensitive to feature scaling (such as Support Vector Machines), min-max normalisation was applied. This step ensures that each feature's values fall within a specific range, preventing any feature from dominating the learning process due to its scale.
\end{itemize}

\subsection{Feature Engineering for Anomaly Detection} One of the critical steps in this research was the extraction of meaningful features from the raw datasets. For both network traffic and process data, feature engineering was used to highlight patterns that might be indicative of cyber attacks. While it’s not feasible to discuss every feature in detail, we focused on those that provided a general context for both network and process behaviour. In particular, features that did not lend themselves well to aggregation, such as specific IP address columns, were excluded. For example, calculating the most frequent IP address for a one-second interval would dilute the critical indicators. An engineer’s workstation might interact with hundreds of devices within a one-second interval, but the presence of a single malicious command-and-control IP would be a much stronger indicator of an attack. Removing features such as IP source and destination allowed the models to detect general trends in traffic and process behaviours rather than specifics. Key features included:

\begin{itemize} \item \textbf{Network traffic features}: Packet sizes, protocol types, connection durations, and byte counts. These features capture patterns in communication that may be altered during a cyber attack. \item \textbf{Process data features}: Sensor readings (e.g., pressure, flow rate) and actuator states (e.g., valve positions). Features such as abnormal sensor spikes or actuator state changes were identified as potential indicators of a physical process under attack. \end{itemize}

\subsection{Machine Learning Models}
Several machine learning models were selected for this study based on their proven performance in intrusion detection and anomaly detection in both \gls{it} and \gls{ot} domains:

\begin{itemize}
    \item \textbf{Random Forests} \cite{breiman2001random}: A robust ensemble method that uses multiple decision trees to make predictions. Random Forests were chosen for their ability to handle mixed data types and their resistance to overfitting. They also provide feature importance rankings, which help in understanding which features contribute most to detecting anomalies.
    \item \textbf{Support Vector Machines (SVM)} \cite{cortes1995svm}: SVMs were selected for their effectiveness in high-dimensional spaces, making them suitable for identifying complex, non-linear patterns in the dataset.
    \item \textbf{K-Nearest Neighbour (kNN)} \cite{kumar2017knn}: A simple, yet effective classifier that predicts the class of a sample based on the majority vote of its nearest neighbours. kNN was used due to its simplicity and effectiveness in identifying outliers or anomalous instances.
    \item \textbf{Neural Networks (NN)} \cite{8386762}: We experimented with both shallow and deep feed-forward neural networks, as well as a 1D-Convolutional Neural Network (CNN) architecture. Neural Networks were chosen for their ability to capture non-linear relationships in the data, making them suitable for detecting subtle anomalies in both network and process data.
    \item \textbf{Ensemble Model with Fuzzy Clustering} \cite{smith1998fuzzy_ann}: To leverage the diverse dataset, we implemented an ensemble model incorporating Fuzzy Clustering. This method allowed the neural network to integrate both explicit information from the original features and implicit patterns discovered through clustering.
\end{itemize}

\subsection{Model Training and Testing}
Each machine learning model was trained and tested using an 80/20 split of the dataset. Cross-validation was employed to ensure the models generalised well to unseen data. The models were trained on three types of datasets:
\begin{enumerate}
    \item \textbf{Network traffic data only}.
    \item \textbf{Process data only}.
    \item \textbf{Combined network and process data}.
\end{enumerate}
This approach allowed us to compare the performance of the models on different data types and assess the benefit of integrating both data streams for anomaly detection.
For hyperparameter tuning, grid search was used to optimise parameters such as the number of trees in Random Forests, the penalty parameter \(C\) in SVM, and the number of neighbours in kNN. For neural networks, different architectures were explored, with various hidden layers and activation functions, and dropout layers were included to prevent overfitting.

\subsection{Data Selection and Aggregation}

To manage the large volume of data and synchronise high-frequency network traffic with process variables, we applied a time-aggregation technique as an essential step in data preparation. Entries were grouped into one-second intervals to align with the sampling rate of the process data, enabling a unified time axis for both network and OT datasets.

Aggregating data requires balancing data integrity with manageable granularity. We used two standard aggregation techniques to simplify the data while retaining essential information: for numerical features, we calculated the mean within each interval to capture representative values; for categorical data, including those transformed via one-hot encoding, we used the mode to maintain the most frequent state within each interval. By reducing millions of raw network packets over a 3-hour 45-minute capture window into one-second summaries, this approach yielded a dataset that remains aligned with the temporal structure needed for anomaly detection while preserving critical details across network and OT events.

The aggregation also helps with noise reduction and mitigates the effects of missing values and outliers. Grouping data by intervals ensures a reliable dataset, providing consistent feature representations for machine learning - across the entire timeline. By structuring network and OT data along a common timeline, this aggregation allows us to test our hypothesis for detecting cross-layer event correlations within the ICS environment. While aggregation into one-second intervals reduces the dataset's size and complexity, it inevitably sacrifices some fine-grained detail. This trade-off between granularity and data manageability is a fine line, but in our case, necessary to enable processing and training.

\subsection{Data Processing and Model Evaluation}

After aggregation, several preprocessing techniques were applied to enhance data quality and optimise the model's detection capabilities. A primary challenge was the class imbalance inherent in cybersecurity datasets like SWaT, where benign instances vastly outnumber attack instances. We addressed this with \gls{smote} \cite{chawla2002smote} to balance the datasets by generating synthetic samples for the minority (attack) class, increasing the likelihood of capturing anomalous behaviour.

\gls{pca} \cite{abdi2010principal} was also used to reduce the data’s dimensionality. This helps simplifying model training while retaining the \enquote{significant} or \textit{principal} features. \gls{pca} is particularly beneficial for high-dimensional datasets, as it mitigates the risk of overfitting while capturing the most impactful data points across both network and process features.

The core focus of this research is to improve the detection capabilities of intrusions across levels in \gls{ics}. Our evaluation centres on \textit{Recall}, 
\[
\text{Recall} = \frac{\text{True Positive}}{\text{True Positive} + \text{False Negative}},
\]
This focus prioritises the model’s ability to detect the majority of attack instances, even at the risk of false positives. In \gls{ics} settings, maximising recall is essential, as missing an attack could lead to severe operational or safety consequences \cite{mishra2019ml_ids}. Additionally, we calculated weighted precision by averaging precision across classes (attack and non attack) weighted by the number of true instances per class, ensuring a balanced evaluation in this unbalanced dataset.

%

\section{Experiment and Discussion}
\subsection{Objective and Metric Choice} \label{section
}

The motivation behind this study is to explore whether incorporating both network and process data can enhance intrusion detection capabilities in \gls{ics}. Given the complex nature of ICS environments, this proof-of-concept research evaluates the potential advantages of using a combined data approach. Specifically, we focus on two key metrics: \textit{weighted precision} and \textit{recall}.

\textit{Weighted precision} addresses the significant imbalance between attack and non-attack instances in ICS datasets, providing a more balanced view of model performance. This metric calculates precision separately for each class (attack and non attack), weighting the precision score by the number of true instances in each class. By doing so, we obtain an indicator of performance that fairly reflects the prevalence of the under-represented attack class. This measure is important for maintaining accuracy in detecting malicious activity without excessively misrepresenting benign instances.

\textit{Recall}, or sensitivity, is equally important, particularly in ICS environments where undetected attacks can have severe operational or safety consequences. This metric emphasizes the model's ability to identify attack instances (true positives) while minimising the likelihood of false negatives. Given our research goal of improving detection for ICS security, recall provides an ideal measure of a model’s capacity to catch attack behaviours accurately across a complex, multi-layered environment. Together, these metrics help to assess whether combining data sources meaningfully contributes to more reliable detection capabilities.

Weighted precision and recall were specifically chosen for this proof-of-concept study due to the unique challenges faced in \gls{ics} environments, where detecting even rare anomalies is crucial. This focus on these metrics helps illustrate the models' potential in minimising false negatives within the constraints of this preliminary setup.

\subsection{Experiment setup}
Each model was trained and tested on three datasets to evaluate the comparative and complementary detection capabilities of network-only, process-only, and combined datasets:

\begin{enumerate}
    \item Network traffic data only: This dataset consists solely of network communication logs (e.g., \texttt{conn.log} and \texttt{cip.log}), capturing information such as packet sizes, connection times, and IP addresses. Network data offers insight into traffic anomalies but is often limited in detecting attacks affecting physical processes.
    \item Process data only: This dataset includes sensor readings and actuator states that directly represent physical process behaviour. Process data is valuable for detecting disruptions in the physical layers of the system, such as sensor spoofing or actuator manipulation, yet may miss network-centric attack patterns.
    \item Combined network and process data: Merging network and process data provides a broader perspective on system state, integrating both traffic and operational insights. By testing the combined dataset, this study seeks to understand if additional features from both data streams improve the model’s ability to detect coordinated or multistage attacks that span both network and process levels.
\end{enumerate}

This approach allows us to observe the distinct contributions of each dataset and to evaluate the benefit of integrating data sources. The results obtained from the singular datasets can be viewed as our \textit{baseline results} and could reflect the capabilities from an IDS in our testbed who only observes the network data, whereas in the \textit{combined network and process data} section, we take a look at at the data combinations we have available and their \enquote{new} capabilities. 

\subsection{Baseline Results}

\begin{figure}[H] \centering \includegraphics[width=0.65\textwidth]{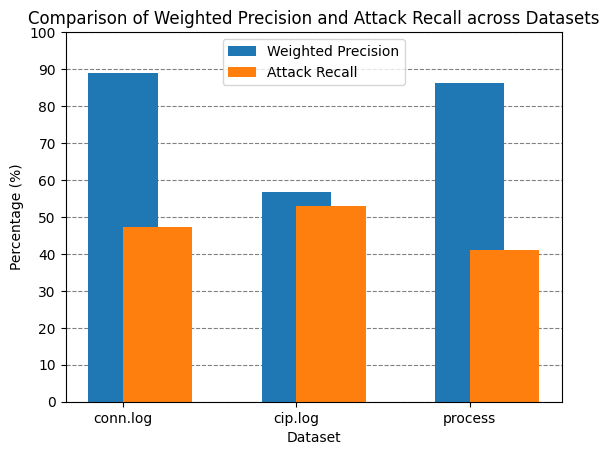} 
	\caption{Histogram showing the distribution of attack recall for baseline models} \label{fig} 
 \end{figure}

The baseline results, summarised in Figure \ref{fig}, show the distribution of attack recall for each dataset without applying \gls{smote}. To not include the \gls{smote} versions in the baseline results is due their overfitting nature, which was observed in preliminary tests where \gls{smote} introduced non-representative synthetic samples, particularly in the \textit{conn.log} and \textit{cip.log} datasets. Among the datasets, \textit{cip.log} showed the weakest performance, likely due to noise present in this dataset.

\subsection{Combined Dataset Results} \label{section}

When combining datasets (\textit{conn.log} + \textit{cip.log}, \textit{conn.log} + \textit{process data}, and \textit{conn.log} + \textit{cip.log} + \textit{process data}), the model performance improved, as illustrated in Figure \ref{fig}. The expanded range of features in the combined datasets appears to enable the models to capture the system’s state more effectively, leading to better detection of anomalies and attack instances across the dataset.

Applying \gls{smote} to the combined datasets further enhanced performance by addressing class imbalances. Unlike in the baseline, where SMOTE introduced overfitting, the richer feature set in the combined datasets seems to allow SMOTE to generate more representative samples along a \textit{broader} feature-range, resulting in more robust learning and improved recall and precision.
\begin{figure}[H] \centering \includegraphics[width=0.85\textwidth]{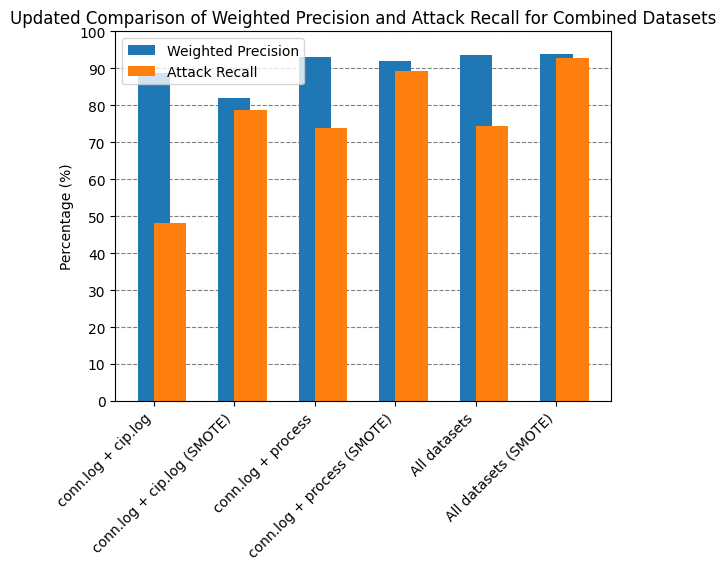} \caption{Histogram showing the distribution of attack recall for combined models} \label{fig
} \end{figure}
The combined datasets, particularly the combination of network traffic and process data, significantly improved model detection performance, as evidenced by increases in recall and precision metrics.

\section{Conclusion}

This research highlights that combining network traffic data with operational process data presents a promising opportunity to enhance and broaden the scope of cyber attack detection in \gls{ics} environments. The results suggest that additional data available in \gls{ot} systems can play a valuable role in improving the depth and scope of anomaly detection. However, these findings are limited to the specific context of the SWaT dataset and the proof-of-concept nature of this study. Future work should extend these investigations across varied \gls{ics} datasets and incorporate more advanced machine learning techniques to better assess the generalisability and robustness of integrated data approaches in detecting complex cyber threats within industrial settings.


\begin{thebibliography}{10}
	\providecommand{\url}[1]{\texttt{#1}}
	\providecommand{\urlprefix}{URL }
	\providecommand{\doi}[1]{https://doi.org/#1}
	
	\bibitem{iec62443}
	Industrial Communication Networks --  part 3-3: System security requirements and security levels,
	\url{https://webstore.iec.ch/publication/6028}.
	
	\bibitem{sklearn_pca}
	{Sklearn Document}, 
	\url{https://scikit-learn.org},
	[Online; accessed Apr. 2024]
	
	\bibitem{abdi2010principal}
	Abdi, H., Williams, L.J.: Principal component analysis. Wiley interdisciplinary
	reviews: computational statistics  \textbf{2}(4),  433--459 (2010)
	
	\bibitem{ackerman2021industrial}
	Ackerman, P.: Industrial Cybersecurity: Efficiently monitor the cybersecurity
	posture of your ICS environment. Packt Publishing Ltd (2021)
	
	\bibitem{alexander2020mitre}
	Alexander, O., Belisle, M., Steele, J.: Mitre att\&ck for industrial control
	systems: Design and philosophy. The MITRE Corporation: Bedford, MA, USA
	\textbf{29} (2020)
	
	\bibitem{breiman2001random}
	Breiman, L.: Random forests. Machine Learning  \textbf{45}(1),  5--32 (2001).
	
	
	\bibitem{chawla2002smote}
	Chawla, N.V., Bowyer, K.W., Hall, L.O., Kegelmeyer, W.P.: Smote: synthetic
	minority over-sampling technique. Journal of artificial intelligence research
	\textbf{16},  321--357 (2002)
	
	\bibitem{cortes1995svm}
	Cortes, C., Vapnik, V.: Support vector networks. Machine Learning
	\textbf{20}(3),  273--297 (1995). 
	
	\bibitem{cyber_report_2023}
	DNV: Cyber priority 2023. Tech. rep., DNV AS, accessed: 1. Desember 2023
	
	\bibitem{galloway2012introduction}
	Galloway, B., Hancke, G.P.: Introduction to industrial control networks. IEEE
	Communications surveys \& tutorials  \textbf{15}(2),  860--880 (2012)
	
	\bibitem{GAMAGE2020102767}
	Gamage, S., Samarabandu, J.: Deep learning methods in network intrusion
	detection: A survey and an objective comparison. Journal of Network and
	Computer Applications  \textbf{169},  102767 (2020).
	
	\bibitem{goh2016}
	Goh, J., Adepu, S., Tan, M., Lee, Z.S.: Anomaly detection in cyber physical
	systems using recurrent neural networks. In: 2017 IEEE 18th International
	Symposium on High Assurance Systems Engineering (HASE). pp. 140--145 (2017)
	
	\bibitem{garcia2009anomaly}
	Garcia-Teodoro, P., Diaz-Verdejo, J., Maci{\'a}-Fern{\'a}ndez, G., V{\'a}zquez,
	E.: Anomaly-based network intrusion detection: Techniques, systems and
	challenges. computers \& security  \textbf{28}(1-2),  18--28 (2009)
	
	\bibitem{hastie2009elements}
	Hastie, T., Tibshirani, R., Friedman, J.: The Elements of Statistical Learning:
	Data Mining, Inference, and Prediction. Springer Series in Statistics,
	Springer New York (2009). 
	
	
	\bibitem{hutchins2011intelligence}
	Hutchins, E.M., Cloppert, M.J., Amin, R.M., et~al.: Intelligence-driven
	computer network defense informed by analysis of adversary campaigns and
	intrusion kill chains. Leading Issues in Information Warfare \& Security
	Research  \textbf{1}(1), ~80 (2011)
	
	\bibitem{jadidi2022automated}
	Jadidi, Z., Foo, E., Hussain, M., Fidge, C.: Automated detection-in-depth in
	industrial control systems. The International Journal of Advanced
	Manufacturing Technology  \textbf{118}(7),  2467--2479 (2022)
	
	
	
	\bibitem{konijn2022multi}
	Konijn, J.P.: Multi-domain Cyber-attack Detection in Industrial Control
	Systems. Master's thesis, University of Twente (2022)
	
	\bibitem{kumar2017knn}
	Kumar, S., Choudhary, G., Kumar, Y.: An analysis of the k-nearest neighbor
	technique for the intrusion detection system. In: 2017 IEEE International
	Conference on Computing and Communication Technologies for Smart Nation
	(IC3TSN). pp. 325--329 (2017). 
	
	 \bibitem{kravchik2018}
	Kravchik, M., Shabtai, A.: Detecting cyberattacks in industrial control systems
	using convolutional neural networks, Proceedings of the 2018 Workshop on Cyber-Physical Systems Security and Privacy, pp. 72-83 (2018)
	
	\bibitem{swat_technical_report}
	iTrust Labs: Swat technical report,
	\url{https://itrust.sutd.edu.sg/itrust-labs\_datasets/dataset\_info/}
	
	\bibitem{liu2019cnn}
	Liu, X., Zhu, J., Li, L.: 1d convolutional neural networks and applications: A
	comprehensive review. Journal of Neural Networks  (2019).
	
	
	\bibitem{mishra2019ml_ids}
	Mishra, P., Varadharajan, V., Tupakula, U., Pilli, E.: A detailed investigation
	and analysis of using machine learning techniques for intrusion detection.
	IEEE Communications Surveys \& Tutorials  \textbf{21}(1),  686--728 (2019).
	
	
	\bibitem{8386762}
	Mishra, P., Varadharajan, V., Tupakula, U., Pilli, E.S.: A detailed
	investigation and analysis of using machine learning techniques for intrusion
	detection. IEEE Communications Surveys and Tutorials  \textbf{21}(1),
	686--728 (2019). 
	
	\bibitem{patcha2007overview}
	Patcha, A., Park, J.M.: An overview of anomaly detection techniques: Existing
	solutions and latest technological trends. Computer networks
	\textbf{51}(12),  3448--3470 (2007)
	
	\bibitem{quinlan1993c45}
	Quinlan, J.R.: {C4.5}: Programs for Machine Learning. Morgan Kaufmann (1993)
	
	\bibitem{shalyga2018anomaly}
	Shalyga, D., Filonov, P., Lavrentyev, A.: Anomaly detection for water treatment
	system based on neural network with automatic architecture optimization
	(2018)
	
	\bibitem{smith1998fuzzy_ann}
	Smith, J.W., Wilson, M.L.: An intelligent hybrid system for pattern
	recognition: Integrating fuzzy c-means clustering with neural networks.
	Neural Networks Journal  (1998). 
	
	\bibitem{stouffer2022guide}
	Stouffer, K., Pease, M., Tang, C., Zimmerman, T., Pillitteri, V., Lightman, S.:
	Guide to operational technology (ot) security. National Institute of
	Standards and Technology: Gaithersburg, MD, USA  (2022).

	
	\bibitem{sullivan2016components}
	Sullivan, D., Luiijf, E., Colbert, E.J.: Components of industrial control
	systems. Springer (2016)
	
	\bibitem{itrust}
	Singapore University of Technology, S.U., (SUTD), D.: {iTrust Labs}.
	\url{https://itrust.sutd.edu.sg/itrust-labs-home/itrust-labs\_swat/},
	accessed: Accessed: 6. Oct. 2023
	
\end{thebibliography}
\end{document}